\documentclass{elsart}

\usepackage{graphicx}
\usepackage{amssymb}

\begin{document}

\begin{frontmatter}

\title{Direct Observation of Broadband Coating Thermal Noise \\
	in a Suspended Interferometer}
	
\date{\today}
\author{Eric D. Black\corauthref{cor1}\thanksref{caltech}}
\ead{blacke@ligo.caltech.edu}
\author{Akira Villar\thanksref{caltech}} 
\author{Kyle Barbary\thanksref{caltech}}
\author{Adam Bushmaker\thanksref{caltech}}
\author{Jay Heefner\thanksref{caltech}}
\author{Seiji Kawamura\thanksref{naoj}}
\author{Fumiko Kawazoe\thanksref{naoj}}
\author{Luca Matone\thanksref{caltech}}
\author{Sharon Meidt\thanksref{caltech}}
\author{Shanti R. Rao\thanksref{caltech}}
\author{Kevin Schulz\thanksref{caltech}}
\author{Michael Zhang\thanksref{caltech}}
\author{Kenneth G. Libbrecht\thanksref{caltech}}
\thanks[caltech]{LIGO Project, California Institute of Technology Mail Code 264-33, Pasadena CA 91125 USA}
\thanks[naoj]{2. National Astronomical Observatory of Japan 2-21-1 Osawa, Tokyo 181-8588 JAPAN}
\corauth[cor1]{Corresponding author}
\begin{abstract}

We have directly observed broadband thermal noise in silica/tantala coatings in a high-sensitivity Fabry-Perot interferometer. Our result agrees well with the prediction based on indirect, ring-down measurements of coating mechanical loss, validating that method as a tool for the development of advanced interferometric gravitational-wave detectors.


\end{abstract}
\end{frontmatter}

\maketitle


There are numerous large-scale interferometer projects around the world aimed at initiating gravitational-wave astronomy, including LIGO \cite{Barish99}, GEO \cite{Luck97}, VIRGO \cite{Caron97}, TAMA \cite{Kawabe97}, and ACIGA \cite{Blair00}. The astrophysical reach of these detectors depends strongly on their strain sensitivity, and even modest reductions in the noise level of a detector can lead to dramatic increases in the number of observed events per year~\cite{Phinney91}. For this reason, it is important not only to reduce the total noise in an interferometric gravitational wave detector to its fundamental limits, but also to understand those fundamental limits and to reduce them as much as possible.

Broadband thermal noise in both the mirror substrates and dielectric coatings is expected to limit the sensitivity of interferometric gravity-wave detectors in their most sensitive frequency bands, in a region that is potentially of the greatest astrophysical interest~\cite{Abramovici92}. Bulk thermal noise can be reduced by using mirror substrates with extremely low mechanical losses, to the point that coating thermal noise is now expected to be a dominant fundamental noise source in advanced detectors~\cite{Levin98,Harry02,Crooks02}.

Since thermal fluctuations are related to mechanical losses via the fluctuation-dissipation theorem~\cite{Callen51,Callen52}, the thermal noise in an interferometer can be calculated if one has sufficient knowledge of the various intrinsic mechanical losses in the system~\cite{Saulson90,Levin98}. Unfortunately, coating thermal noise is considerably more difficult to model than bulk thermal noise, owing to the complex, multi-layer structure of dielectric coatings. Furthermore, it has been shown that the mechanical and thermal properties of thin films in general can differ markedly from the bulk properties of identical materials~\cite{Wu93}.

To investigate the fundamental measurement limits imposed by coating and bulk thermal noise, we constructed a small-scale suspended interferometer to directly measure thermal noise. We used fused-silica test masses and low-loss $\mbox{SiO}_2/\mbox{Ta}_2\mbox{O}_5$ coatings with a fairly large laser spot size, so the thermal noise was as low as practical, and thus pertinent to the investigation of thermal noise in gravitational wave detectors.

\section{The instrument}

\begin{center}
\begin{figure}
\includegraphics{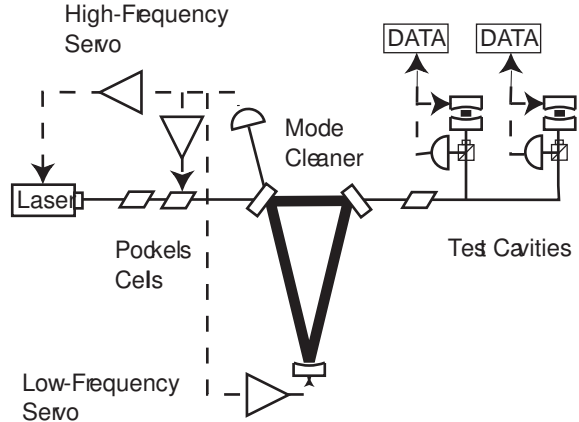}
\caption{\label{fig:layout} Schematic of the Thermal Noise Interferometer (TNI). All of the mirrors in the mode cleaner and arm cavities were suspended under vacuum. The mode cleaner provided both spatial filtering and a frequency stabilization reference. We locked the test cavities to the resulting beam, and we recorded the data directly from the test cavities' error signals. Common-mode rejection was implemented by taking the real-time difference between the two data streams.}
\end{figure}
\end{center}

Figure~\ref{fig:layout} shows a schematic of the interferometer. Two test cavities with identical lengths and finesses were made from four identical, independent, suspended mirrors. Three other suspended mirrors made up a mode cleaner, which provided both spatial filtering and, through an active feedback system, frequency stabilization for the laser. The test cavities were locked to the resulting filtered and stabilized beam by actuating on their output (end) mirrors. Residual laser frequency noise was partially removed by differencing matched data streams from the two cavities.  


Each suspended optic was supported by a single loop of fine steel music wire and actuated on by means of a magnet-coil system, with the magnets being attached to the backs of the mirrors. Each suspended optic was also actively damped using analog feedback electronics. Test mirrors were made from high-purity, synthetic fused silica (Corning 7980). All optics were superpolished on their faces for optical performance and polished on their barrels to improve mechanical $Q$'s. High-reflectivity coatings were multilayer $\mbox{Si}\mbox{O}_2/\mbox{Ta}_2\mbox{O}_5$ dielectric stacks provided by Research Electro-Optics~\cite{REO}.
\begin{center}
\begin{figure}
\includegraphics{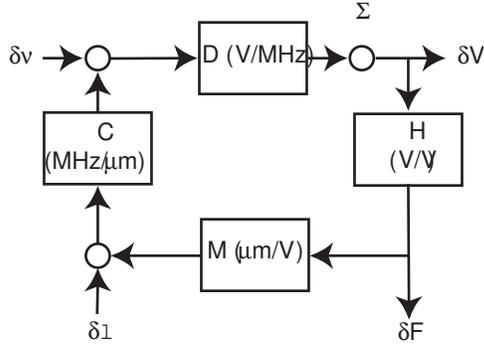}
\caption{\label{fig:control} Block diagram of the servo used for locking each arm cavity. Here $\delta \ell$ is the cavity length noise, and $\delta V$ is the measured voltage fluctuations that result. Knowledge of the transfer functions $D$, $H$, $M$, and $C$~was used to calculate $\delta \ell$ from the measured $\delta V$.}
\end{figure}
\end{center}

The test cavities were locked to the stabilized laser beam using the Pound-Drever-Hall method~\cite{Drever83,Black01}. Figure~\ref{fig:control} shows a block diagram of the servo used for locking the arm cavities and for acquiring data. Here $D$ represents the Pound-Drever-Hall discriminant, $H$ the electronic transfer function of the servo filter, $M$ the transfer function of the actuation system, and $C=\nu / L$ is a conversion factor between the length of the cavity $L$ and the laser frequency $\nu$. The equivalent length noise of the cavities was obtained from the measured voltage $\delta V$ by
\[
\delta \ell = \frac{1+DHMC}{DC} \delta V.
\]

We had to know the frequency-dependent transfer functions of the blocks in Figure~\ref{fig:control} in order to convert the measured voltage noise $\delta V$ into an equivalent length noise $\delta \ell$. The electronic transfer function $H$ was specified in the design of the instrument and verified by direct measurements. The mirror response $M$ was measured two independent ways: First, we constructed a Michelson interferometer with the suspended mirror forming the end mirror of one arm. By driving the mirror through multiple fringes, we were able to calibrate its response. Second, we drove the PZT input of the laser to introduce a known frequency fluctuation $\delta \nu$ into the beam through the mode cleaner with the system in lock. We then measured the feedback voltage $\delta F$ at frequencies well below the unity-gain frequency, which allowed us to determine $M$. We calibrated $\delta \nu$ using a fixed-length Fabry-Perot cavity, and our measured PZT response agreed with the manufacturer's specifications. Both methods of calibration gave the same value for $M$, within experimental uncertainties.



The Pound-Drever-Hall discriminant $D$ was also measured two different ways. First, we measured the slope of the error signal at the resonance point, using the sidebands as a frequency reference, as the system was swept through resonance with no feedback engaged. Second, we locked the arm cavities and measured the total open-loop transfer function $DHMC$ using a summing junction ($\Sigma$ in Figure~\ref{fig:control}). Using known values for $H$, $M$, and $C$, we then fit a theoretical prediction of $DHMC$ to this measurement, with $D$ as an adjustable parameter. Both determinations of $D$ agreed to within experimental uncertainties. This open-loop transfer function measurement and calibration were performed separately each time the interferometer was locked.

\begin{center}
\begin{figure*}
\resizebox{\textwidth}{!}
	{\includegraphics{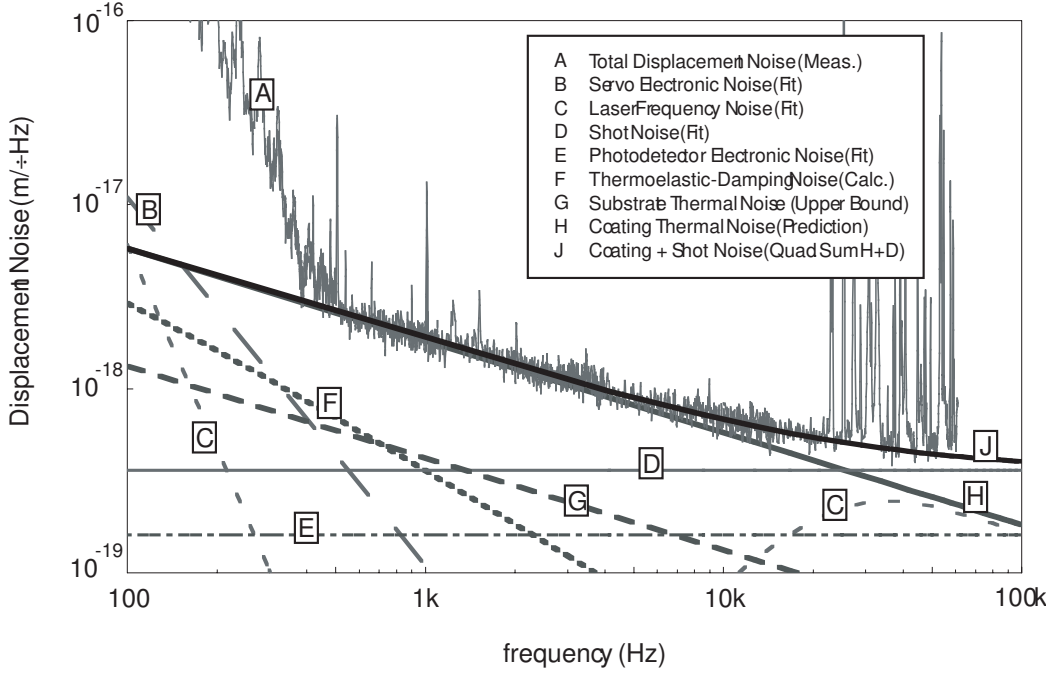}}
\caption{\label{fig:result} Displacement noise spectrum of the Thermal Noise Interferometer, along with contributions from both technical and fundamental noise sources. Technical noise source curves (servo and photodetector electronic noise, laser frequency noise) are fits to independent measurements. Of the fundamental noise curves, coating thermal noise and shot noise are fits to measurements, whereas substrate thermal noise and thermoelastic-damping noise are calculations.} 
\end{figure*}
\end{center}

We performed two additional tests on the data by checking the scaling of the noise floor with laser power and with the modulation voltage applied to the resonant Pockels cell just before the test cavities. In both cases, we found that the total noise in the instrument was independent of these quantities at frequencies where thermal noise dominates. Noise that originated before the cavities was measured at the input of the mirror actuation system ($M$ in Figure~\ref{fig:control}) and is labeled "Servo Electronic Noise" in Figure~\ref{fig:result}. This noise was well below the total noise curve at all frequencies. Thus we verified that the noise originated in the test cavities and not anywhere else in the control loop or measurement signal path. 

\section{Results}

Figure~\ref{fig:result} shows the total displacement noise of the interferometer, with common-mode rejection implemented, along with contributions from various noise sources. The theory curve (J) is the sum, in quadrature, of the coating thermal noise in all four mirrors in the arm cavities and the measured constant shot noise. The theoretical model we use takes into account the different Young's modulus and Poisson's ratio of the coating and substrate, and it admits the possibility that the coating mechanical loss angle might be different for strains parallel and perpendicular to the substrate-coating interface. For a coating with thickness $d$, Young's modulus $E_c$, and Poisson's ratio $\sigma_c$, the thermal noise is~\cite{Harry02}
\begin{eqnarray}
\delta \ell^2_{coat} (f) &=& \frac{2 k_B T}{\pi^{3/2} f} \frac{1 - \sigma^2}{E w_0} \left\{ \frac{1}{\sqrt{\pi}} \frac{d}{w_0} \frac{1}{E E_c (1-\sigma_c^2)(1-\sigma^2)} \right. \nonumber \\
& \times & \left[ E_c^2 (1+\sigma)^2(1-2 \sigma)^2 \phi_{\parallel} \right. \nonumber \\
&+& E E_c \sigma_c (1+\sigma)(1+\sigma_c)(1-2\sigma)(\phi_{\parallel} - \phi_{\perp}) \nonumber \\
&+& \left. \left. E^2 (1+\sigma_c)^2(1-2 \sigma_c) \phi_{\perp} \right] \right\}
\label{eq:full-coating}
\end{eqnarray}
where $\phi_{\parallel}$ and $\phi_{\perp}$ are the coating's mechanical loss angles for strains parallel and perpendicular to the substrate-coating interface, respectively. ($E$ and $\sigma$ are the Young's modulus and Poisson's ratio for the substrate.) 

If we assume the previously-reported value for $\phi_{\parallel}$ of $2.7 \times 10^{-4}$~\cite{Penn03}, we find that the perpendicular loss angle is the same as the parallel loss angle for $\mbox{SiO}_2/\mbox{Ta}_2\mbox{O}_5$ coatings. It would be premature, however, to conclude that the loss is definitely isotropic, because of the large error bars imposed on $\phi_{\perp}$ by the uncertainty in $\phi_{\parallel}$ and by the relatively weak dependence of $\delta \ell$ on $\phi_{\perp}$.

Note that, if we assume that the coating Young's modulus and Poisson's ratio are the same as those of the substrate, and that the mechanical loss of the coating is the same for strains parallel and perpendicular to the substrate-coating interface, the coating thermal noise formula takes on a particularly simple form~\cite{Harry02}, 
	\begin{equation}
	\label{eq:isotropic-coating}
	\delta \ell^2_{coat} (f) \approx \frac{2}{\pi^{3/2}} \frac{k_B T}{f} \frac{1 - \sigma^2}{E w_0} \left\{ \frac{2}{\sqrt{\pi}} \frac{d}{w_0} \left( \frac{1 - 2 \sigma}{1 - \sigma} \right) \phi_{\parallel} \right\}.
	\end{equation}
Using the value for the loss angle $\phi_{\parallel}$ obtained from the ringdown technique~\cite{Penn03}, we have an unambiguous prediction of the coating thermal noise that we can compare with our direct measurement. This prediction yields a theoretical curve that is within a few percent of the fit we performed using Equation~\ref{eq:full-coating} and thus agrees well with our observed noise floor.

The parameters used in our analysis were
\begin{center}
\begin{tabular}{lcc} 
Substrate Young's Modulus: & $E$ & $7.0 \times 10^{10} N/m^2$~\cite{Harry02} \\
Coating Young's Modulus: & $E_c$ & $11.0 \times 10^{10} N/m^2$~\cite{Harry02} \\
Substrate Poisson's Ratio: & $\sigma$ & $0.17$~\cite{Braginsky99-2} \\
Coating Poisson's Ratio: & $\sigma_{c}$ & $0.20$~\cite{Harry04-T} \\
Coating thickness: & $d$ & $4.26 \mu m$~\cite{REO} \\
Laser spot radius: & $w_0$ & $160 \mu m$ \\
Parallel loss angle: & $\phi_{\parallel}$ & $(2.7 \pm 0.7) \times 10^{-4}$~\cite{Penn03} \\
Perpendicular loss angle: & $\phi_{\perp}$ & $(2.7 \pm 2.2) \times 10^{-4}$ (fit) \\ 
\end{tabular}
\end{center}

The uncertainty in $\phi_{\perp}$ due to the fit is approximately ten percent. However, since the noise curve is roughly three times more sensitive to $\phi_{\parallel}$ than it is to $\phi_{\perp}$, the modest uncertainty in $\phi_{\parallel}$ results in a fairly large uncertainty in $\phi_{\perp}$.

The parallel loss angle $\phi_{\parallel}$ reported in Reference~\cite{Penn03} is for coatings made by a different manufacturer from the ones on our mirrors, but one of the conclusions of that paper was that most of the mechanical loss in a $\mbox{SiO}_2/\mbox{Ta}_2\mbox{O}_5$ coating is intrinsic to the tantala layers. Our coatings nominally have the same chemical content and physical structure as those reported in Reference~\cite{Penn03}, so we assume that their results apply to our coatings.

It has been suspected for some time that the initial estimates of LIGO's thermal noise floor~\cite{Abramovici92} were high, because a viscous-damping model was assumed for the loss mechanism, as opposed to the structural-damping model elaborated after the initial design studies~\cite{Saulson90,Levin98,Numata03}. Because the expected event rate scales so strongly with the sensitivity level in a gravitational wave detector, even a small reduction in the fundamental noise floor can translate into a large increase in the event rate. If, as our results indicate, the noise floor of LIGO is dominated by structural-damping-mediated coating thermal noise with a coating loss of $2.7 \times 10^{-4}$, the effective range for binary neutron star inspiral detection of this first-generation instrument could be as much as $28 \mbox{Mpc}$, almost twice the original estimate of $15 \mbox{Mpc}$~\cite{Adhikari04}. For an isotropic distribution of sources, this would mean an almost eightfold increase in the rate of observed events.

Greater are the implications for advanced detector development. Currently there is a substantial effort to develop a coating that will exhibit lower levels of mechanical noise than the $\mbox{SiO}_2/\mbox{Ta}_2\mbox{O}_5$ coatings used in the existing LIGO mirrors. The agreement, both qualitative and quantitative, between the indirect, ringdown measurements of coating loss and our direct observation of coating thermal noise gives us confidence in our understanding of this noise source in a real, high-sensitivity interferometer, allowing the prediction, and even design, of the noise floor of an advanced detector.



\section{Acknowledgments} 

We acknowledge useful discussions with Stan Whitcomb and Gregg Harry, and electronics assistance from Rich Abbott and Flavio Nocera. This work was supported by the NSF under grant number PHY01-07417.

\bibliographystyle{elsart-num}
\bibliography{/Users/eric/Documents/Bibliographies/astrophysics,/Users/eric/Documents/Bibliographies/thermal-noise-story,/Users/eric/Documents/Bibliographies/Photothermal}

\end{document}